\documentstyle[aps,pra,epsf,floats,floats]{revtex}

\newcommand{\be}{\begin{equation}}
\newcommand{\ee}{\end{equation}}
\newcommand{\bear}{\begin{eqnarray}}
\newcommand{\eear}{\end{eqnarray}}

\begin{document}

\ifpreprintsty \else
\twocolumn[\hsize\textwidth\columnwidth\hsize\csname@twocolumnfalse%
\endcsname \fi

\draft 

\title{Patterns of consumption in  a discrete choice model
with  asymmetric  interactions}

\author{Giulia~Iori}
\address{Department of Mathematics, King's College, University of  London\\ 
The Strand, London, WC2R 2LS, UK.}
\author{Vassilis~Koulovassilopoulos}            
\address{Business School, South Bank University, 103 Borough Road,
London, SE1 0AA, UK.}

\date{\today}
\maketitle

\begin{abstract}
We study the  consumption behaviour of an asymmetric  network of heterogeneous 
agents  in the framework of discrete choice models 
with stochastic decision rules.
We assume that the interactions among agents are  uniquely specified by
their ``social distance'' and consumption  is driven by peering,
distinction and aspiration effects. The utility of each agent is positively or
negatively affected by the choices of other agents and consumption is
driven by peering, imitation and distinction effects.  
The dynamical properties of  the model are explored, by numerical
simulations, using three different evolution algorithms with: {\it
parallel, sequential} and {\it random-sequential} updating rules.
We analyze  the long-time behaviour of the system which,
given the asymmetric nature of the interactions,  can either converge
into  a fixed point or a periodic attractor. 
We  discuss the role of symmetric versus asymmetric contributions to the
utility function and also that of idiosyncratic preferences,  costs and
memory in the  consumption decision of the agents.

\end{abstract}

\pacs{} 

\ifpreprintsty \else
] \fi              

\section{Introduction}

A great body of research has been devoted to the effects that direct
interactions  among consumers or firms  have on macroeconomics
variables (Brock 1995, Brock and Durlauf 1995, Aoki 1996,  Axelrod 1997,
Albin 1998, Chwe 2000). 
Direct interactions among economic agents, usually referred to
as social interactions (as opposed to market mediated interactions) are  
meant to capture how the decision of each individual is influenced by
the choice of others in his reference group.   
Direct interaction models can apply to coordination problems in general,
ranging from the emergence of collective political actions and the
development of fads and conventions to the explanation of speculative
bubbles in financial markets and the dynamics of market penetration and
diffusion of technological innovations.

Different alternatives have been considered in the 
literature: global interactions (Brock and Durlauf 1995), 
where each individual tends to conform to the average behaviour of the
entire population, as well as local interactions, where each individual
has an incentive to conform to a specified group of neighbours (F\"ollmer
1974, Durlauf 1993, Blume 1993, Corneo 1994, Morris 2000). 
This last case has recently  gained
interest in economics, following the observation that network
externalities are often localized.  
A stochastic interaction picture has also been considered.  This can 
be implemented by considering either fixed, exogenously determined,
random communication links between any pair of agents or, by taking
time-dependent links and letting the neighbouring composition evolve in
a self-organized way (Benabou 1996, Durlauf  1996). In this case, agents
would be able to form new alliances according to some fitness
maximization scheme.   
Moreover, links among agents can be of varying strength, often positive
and negative (Axelrod 1997, Galam 2000), to account for the different
externalities an agent receives from the behaviour of the other agents,
depending not only on where they are but also on who they are.
 
Nonetheless, in the literature the attention has been mainly focused on
the case of positive, pairwise symmetric, spillover, i.e. the case where
the payoff of a particular action increases when others behave similarly. 
In this context, it has been shown that the evolution is diffusive: even
in the case of heterogenous agents, social interactions create
conformity in behaviour or polarized group behaviour without relying on
the presence of correlated characteristics among members of the same
group. 
While models with symmetric interactions have given numerous insights
in a variety of contexts, from a sociological point of view the
constraint of symmetric interactions is unsatisfactory. Two agents do
not need to influence one another in the same manner. Therefore it is
natural to investigate models with asymmetric couplings between agents. 

Non-symmetric pairwise interactions, although common in the study of
neural networks and other biological systems (Kauffman 1969, see also 
M\"uller et al. 1995 for a review), have only recently been introduced in
economics (Akerlof 1997, Kirman 1997, Samuelson 1997).  
In a recent work, Cowan, et al. (1998), introduced a model of consumption
behaviour, herefrom called the CCS model, where the utility of an individual
agent is positively or negatively affected by the choices of other
agents and consumption is driven by peering, imitation and distinction
effects.  
In the CCS model, the microeconomic agents have pairwise interactions
which are specified by a function of a single parameter, their ``social
distance'' such as, for example, differences is wealth (where the 
wealth can be a random variable). 
Consumers are ordered according to their social status
and are affected by the behaviour of other agents depending on their
relative location on the spectrum. Agents wish to distinguish themselves
from those who are below and emulate their peers and those who are above
in the social spectrum. The interplay between aspiration  and
distinction effects can generate consumption waves, which propagate
through the system. 

The CCS model has been analyzed in the framework of random utility
discrete choice models, extending the literature there by combining both local
and global externality effects.  
Discrete choice models start from the assumption that each agent faces a 
set of mutually exclusive alternatives, and chooses the one that 
yields greatest utility (see Anderson et al. (1992) for a review).
Two families of models have been introduced to analyze the choice
process in a probabilistic setting.  
The first family of models assumes that the decision rule is
deterministic but the utility is stochastic.  
The idea behind this assumption is that even though individual behaviour
might be deterministic, the modeler can only imperfectly observe the
factors that influence individual choice and only has an imperfect
knowledge of the utility function of each  agent. 
Models in the second family assume that the utility is deterministic but
the choice process is stochastic. These models capture the idea of
bounded rationality of economic agents (Sargent 1995). 
Even if utility is deterministic, individuals might make an error in
evaluating the importance of one or another characteristic associated
with a certain alternative and do not necessarily select what is best
for them. The CCS model lies in the first category, namely that of
random utility models with interacting agents.  
In this paper we reformulate the CCS model and adopt the alternative
description, that of the choice as a stochastic process with the utility
being deterministic. 

Discrete choice models have been analyzed using the techniques of
statistical mechanics. In the case of symmetric interactions, the
equilibrium condition can be expressed in terms of the Boltzman
distribution. This is no more the case in models with
asymmetric interactions. 
As a consequence, the long time behaviour of the system has to be
calculated by solving the dynamical problem (which in most cases is
not possible analytically) and cannot be evaluated by equilibrium
ensemble averages.  
We use numerical simulations with Glauber dynamics (Glauber 1963) 
to explore,  the dynamical properties of the  model.
We implement three different evolution algorithms with: {\it parallel,
sequential} and {\it random-sequential} updating rules, depending on the
order on which individual agents update their decision.
We first focus on  the deterministic limit and study the attractors of
the model, which determine the steady state, long-time behaviour of the
consumption behaviour.  
Depending on the evolution algorithm as well as the degree of the asymmetry
 the attractors can be either fixed points or limit
cycles. We then introduce noise in the system and study how this affects
the  dynamics of consumption.
Eventually, extending the analysis of CCS we discuss the
role of costs and memory in the consumption decision of the agents and 
consider different scenarios for the connectivity among the economic agents.

In section \ref{DCM} we set the general framework by reviewing briefly
the theory of discrete choice models in a stochastic environment. In
section \ref{Model} we present our  model in terms of the
interactions and the possible evolution mechanisms. Section
\ref{results} contains results from our simulations while section 
\ref{conclusions} conclude.

\section{Discrete Choice Models} 
\label{DCM}

Discrete choice models can be formalized (Brock and Durlauf 1995,
Durlauf 1997) by considering a population of $N$ individuals, where each
individual $i$ chooses $S_i$ with support $(-1, 1)$. The set of all
possible sets of actions by the population, denoted by $\Omega$,
consists of all N-tuples $\tilde S = (S_1, \ldots, S_N)$.

In the following  we  briefly review the basic ideas of both stochastic
utility and stochastic decision rule  approaches.

\subsection{Stochastic utility models} 

This problem has been formulated initially in the case of non interacting
 agents (Anderson et al. 1992) and has been subsequently
 generalized  by Brock and Durlauf (1995) to the case 
with social interactions.

Individual utility $U_i(S_i)$ consists of two components, a
deterministic term plus a random component: 
\be
U_i(S_i) = V_i(S_i) + \epsilon_i(S_i)
\ee
The probability that an individual chooses $S_i = 1$ is  given by
\be
P(S_i = 1) = P(V_i(1) + \epsilon_i(1) > V_i(-1) + \epsilon_i(-1))
\ee
or
\be
P(S_i=1) = P( \epsilon_i(-1) - \epsilon_i(1) < V_i(1) - V_i(-1)) 
\ee
To solve the problem one needs to make some assumption about the distribution 
of the random terms.
The  random  disturbances are assumed  independent and identically
distributed across agents, and are known  at the time agents  take their
decision. 
 Let $f(\epsilon)$  be the distribution of
 $\epsilon = \epsilon(1) - \epsilon(-1)$.
Then 
\be
P(S_i=1) = \int_{-\infty}^{V_i(-1) - V_i(1)} f(x) dx
\ee
If $f(x)$ is logistically distributed with zero 
mean and variance $\mu$ 
\be
P(\epsilon \le z) = \frac{1}{1+exp(-z/\mu)} 
\ee
then
\be
P(S_i=1) = \frac{e^{U_i(1)/\mu}}{e^{U_i(1)/\mu} + e^{U_i(-1)/\mu}}
\ee
The generalization to multiple choices is possible if 
 the double exponential  distribution is assumed for the noise
\be
P(\epsilon_i \le z) = exp(-(exp(-z/\mu + \gamma)))  
\ee
where $\gamma$ is the Euler's constant ($\gamma \sim 0.5772$).
In this case
\be
P(S_i = \nu_i ) = \frac{e^{V_i(\nu_i)/\mu}}
	{\sum_{j=1}^n  e^{V_j(\nu_j)/\mu}}
\ee

This  problem has been generalized by Brock and Durlauf, when
social externalities affect agents' decisions, by taking
 into account an extra term in the utility function:
\be
	U_i(S_i)  = u_i(S_i) + s_i(S_i, \mu^e(\tilde S_{-i})) +
	  \epsilon_i(S_i) 
\ee
where $u_i(S_i)$ represent agent's $i$ deterministic  private utility, 
 $s_i(S_i, \mu^e(\tilde S_{-i}))$ represent his/her deterministic   social
 utility,  and $\epsilon(S_i)$ represents a random private  utility.

The term $\mu^e(\tilde S_{-i})$ denotes the conditional probability
 measure agent $i$ places on the choices of others at the time of making
 his own decision.
Brock and Durlauf write the  social utility  as 
\be
 s_i(S_i, \mu^e( \tilde S_{-i})) = 
	\sum_{j \in G_i} J_{ij} S_i E_i[S_j] 
\ee
where $G_i$ is the reference group of agent $i$ and 
$E_i[.]$ represent the conditional expectation operator associated
 with agent $i$'s beliefs.  $J_{ij}$
 represent the interaction weight  which relates $i$'s 
choice to  $j$'s choice. 
The $J_{ij}$ are  chosen 
equal to 1 if a link between  a pair  $(i,j)$ exist and zero eitherways. 
Within this framework, choosing a particular realization
 of $J_{ij}$, the interactions among agents are completely specified.

The private utility $u_i(S_i)$ is assumed to depend
linearly on $S_i$: 
\be
  u_i(S_i) = h_i  S_i
\ee
where the $h_i$ can be chosen the same  for all agents or can have different
values for different $i$ if we assume heterogeneous agents.
 Note that $h_i$ plays the role of an external field, affecting the
decision of agent $i$. Analogously, the social utility term can be
interpreted as an internal field generated  by the agents themselves.
Defining 
\be
h_i^t = h_i S_i + \frac{1}{N}\sum_{j \in  G_i} J_{ij} E_i[S_j]
\ee
the deterministic component of the utility becomes
\be
 V_i(S_i) = h_i^t S_i.
\ee
Note we have rescaled the interactions $J_{ij}$, dividing by $N$, in order to keep $U_i$ finite as $N \rightarrow \infty$.

If the $\epsilon(S_i)$ are independent across agents, 
the joint probability measure over all agents choices  equals:
\be
P(\tilde S) = \prod_{i=1}^N P_i(S_i)
\ee
and we can rewrite eq.(9) as 
\be
P(\tilde S) = \frac{\exp{(\sum_{i=1}^N V_i(\nu_i)/\mu)}}
	{\sum_{\nu_1} ... \sum_{\nu_N} 
	\exp( \sum_{i=1}^N  V_i(\nu_i)/\mu)}
\ee
	
\subsection{Stochastic decision rules models}

Following the observation that adjustments to the behaviour of
economic agents are often made at discrete points in time and are
of finite magnitude, one can  use jump Markov processes to model
the evolutionary dynamics of a large collection of interacting microeconomic
agents (Aoki 1996). 
Interactions of microeconomic units can then be specified in terms of
transition probabilities of Markov chains (discrete Markov process 
with finite state-space).
The initial condition and the state transition probability completely
characterize the time evolution of a discrete-time Markov chain.
The time evolution of the probabilities of states in terms of transition
rates and the state's occupancy probability is given by the master
equation: 
\be
\frac{\partial P(x',t)}{\partial t} = \sum_{x \neq x'} P(x,t)w(x'|x,t) -
\sum_{x \neq x'} P(x',t)w(x|x',t)  \label{master}
\ee
where $x, x^\prime$ denote state space variables. From this equation one
sees that for the stationary or equilibrium probability to exist
the (full) balance condition should be satisfied:  
\be
\sum_{x \neq x'} P_e(x)w(x'|x) =
\sum_{x \neq x'} P_e(x')w(x|x') 
\ee
If moreover the probability flow balances for every pair of states
 $(x,y)$, i.e. the detailed balance condition holds
\be
 P_e(x)w(y|x) = P_e(y)w(x|y) \label{detailed-balance}
\ee
it can be shown that the equilibrium distribution is path independent.
More precisely, denoting by $\Omega$ the space state of a Markov chain,
if we assume that the Markov chain is ergodic\footnote{More precisely,
the chain should be irreducible, aperiodic and positive (see, for
example, Hammersley and Handscomb 1975)},
any state $x_i \in  \Omega$ can be reached from an initial state $x_0$
through a sequence of intermediate states $x_1, x_2,..., x_{i-1}$ so
that   
\be
 P_e(x_i) = P_e(x_0) \prod_{k=0}^{i-1}w(x_{k+1}|x_k)/w(x_k|x_{k+1}) .
\ee
If the detailed balance condition holds it can be shown that
\be
 P_e(x_i) = C w(x_0|x_i)/w(x_i|x_0) 
\label{nopath} 
\ee
In other words $P_e(x)$ is a Gibbs distribution
\be
 P_e(x) = \frac{\exp ( V(x))}{\sum_{k \in \Omega} \exp(V(k))}
\label{gibbs} 
\ee
where
\be
V(x_i) - V(x_0) = \log (w(x_i | x_0)/w(x_0 | x_i) ).
\ee
$V(x)$, depending  only  on the state $x_i, $ is a
potential. It follows that
any dynamical process, as long as it satisfies the 
detailed balance condition with the same function $P_e$, will 
 reach the same asymptotic equilibrium distribution of
states and, from eq.(\ref{nopath}), 
 that the equilibrium probability distribution
 is independent of the dynamical trajectory in the
configuration space. 

Given a potential $V(x)$ which we want to maximize
the problem eventually becomes  
to find  the appropriate transition probabilities 
which satisfy the detailed balance condition with $P_e(x)$ specified by
eq.(\ref{gibbs}).
The Glauber dynamics, or heat-bath algorithm (Glauber 1963), serves this
purpose. 
In our formulation we want  the potential $V(x)$ to be the total
utility $U(\tilde S)$ of the system.
 We assume here that the utility is the same as the one given
 in the previous section except for the random component  which 
is now missing:
\begin{equation}
U_i(S_i) = \frac{1}{N} \sum_{j=1}^N  J_{ij} S_i E_i[S_j] + h_i  S_i
\label{utility} 
\end{equation}
The total utility $U(\tilde S)$ of the system in the configuration 
$\tilde S$ is then
\be
U(\tilde S) = \sum_{i=1}^N U_i(S_i) \label{tot_util}
\ee

Accordingly, to the heat-bath algorithm the probability of a agent $i$
to take a value $S_i$, where $S_i$  can only be +1 or  -1, is
\be
Pr(S_i) = \frac{\exp(\beta h_i^t  S_i)}{\exp(\beta h_i^t ) 
        + \exp(-\beta h_i^t)}
\label{HeatBath}
\ee
This can be interpreted  as the choice process of our agents
not being  entirely deterministic. 
Indeed, we are assuming that there is a ``noise'' element 
 in the agents' decision 
represented by a Glauber dynamics 
 with a thermal agitation characterized 
by a temperature $T=1/\beta$ which is held fixed.

It is easy to show that the updating  process described by
eq.~(\ref{HeatBath})  obeys the detailed balance condition with $P_e =
\exp(- \beta U(\tilde S))$ if the $J_{ij}$ are symmetric, i.e. $J_{ij} = 
J_{ji}$. 
 Following Amit (1989), we  note first that the 
probability for a agent $i$ to go to a state $S_i$ only depends on the
 final state of that agent and the states of all other agent, who create 
the local field. Therefore the probability for the system to go from 
a configuration $\tilde S_J$, in which the agent $i$ is in state $S_i$ to a 
configuration  $\tilde S_I$, in which that agent is in state $-S_i$, 
is $Pr(-S_i)/N$
and the probability to go from $\tilde S_I$ to $\tilde S_J$ is $Pr(S_i)/N$
(the denominator expresses the probability to pick up that specific agent). 
Hence
\bear
\frac{w(\tilde S_I|\tilde S_J)}{w(\tilde S_J|\tilde S_I)} &  = &
 \frac{Pr(S_i)}{Pr(-S_i)} = 
\frac{\exp(\beta h_i^t S_i)}{\exp(-\beta h_i^t S_i)} 
 \nonumber \\
 & = & \exp( \beta U(\tilde S_I) + U(-\beta \tilde S_J))
\eear
from where the detailed balance condition holds and hence
\be
P_e \sim  \exp( \beta U(\tilde S)).
\label{gibbs2} 
\ee

Nonetheless if  there are asymmetric interactions, i.e. 
$J_{ij} \neq J_{ji}$ the detailed balance condition, 
eq.~(\ref{detailed-balance}) is no longer valid, and the long time 
evolution of the system does not necessarily converge to the Gibbs
equilibrium distribution defined in  eq.~(\ref{gibbs2}). 
Even though nothing can be said, a priori, in this case about the
equilibrium distribution, the heat-bath algorithm still  gives 
a prescription on how to introduce a dynamics which, by maximizing 
$U(\tilde S)$, leads to an equilibrium state.
Moreover, given that the contribution of any asymmetric
component of the interactions to the total utility $U(\tilde S)$
is zero when calculating the double sum, $U(\tilde S)$ would not be a
Lyapunov function in this case. 
We remind that in the absence of a Lyapunov function the system is not
forced to approached  a stable attractor state and limit cycles can
occur as well as fixed points.

Note that in eq.~(\ref{HeatBath}) we have not specified the order of
updating. 
Two versions of the dynamics have become popular: the first assumes
that all agents  update their state simultaneously at every discrete time
step $t_i$. 
The state of the other agents is in this case considered to be the one in the
time interval $t_{i-1}$.
This type of dynamics is called {\it synchronous} or {\it parallel}.
The second kind of dynamics is the {\it asynchronous} or {\it
sequential} in which the state of each agent  is updated one by one.
In this case  every agent coming up for a decision has full information 
about the state of the other agents that have been updated before him.

In the following sub-section  we focus  on the dynamics at $T=0$ (the
deterministic limit) and review  the main results concerning the  nature
of the attractors when using  different updating rules.

\subsection{Dynamics}
\label{evolution}

\bigskip 
(a) {\it Parallel (or synchronous) dynamics:}
At each time step $t$, all agents re-evaluate  {\it simultaneously} their
consumption decision, relative to that taken one step before, on the
basis of the utility $U_i(t)$ they receive, according to:
\begin{equation}
S_i(t + 1) = \mbox{\rm sgn} (h_i^t) =
      \mbox{\rm sgn} \left[ \sum_{j=1}^{N} J_{ij} S_j(t) + h_i \right]
\label{parallel}
\end{equation}
where we have implicitly absorbed the $1/N$ factor in the definition of
$J_{ij}$ and introduced for later convenience the notation:
\begin{equation}
 h_i^t= h_i + \sum_{j\neq i} J_{ij} S_j  .
\label{hit}
\end{equation}

(b) {\it Sequential dynamics:}
This is the case of asynchronous updating in which the state of each
agent  is updated one by one in a serial manner, according to:
\be
S_i(t+1) = \mbox{\rm sgn} \left[\sum_{j < i} J_{ij} S_j(t + 1) +
	\sum_{j > i} J_{ij} S_j(t) + h_i \right]
\label{sequential} 
\ee

(c) {\it Random sequential (Glauber) dynamics:} 
This is another case of asynchronous updating scheme, similar to the
previous case, with the difference though that the order of updating is
chosen randomly.

For both synchronous and asynchronous dynamics, symmetric couplings 
$J_{ij} = - J_{ji}$ is a
sufficient condition for the existence of detailed balance. Nonetheless  
the form of the asymptotic distribution differs in the two dynamics.
In the following, we briefly review the properties of the three dynamical
 rules.
 One can draw a number of general statements about the nature of
fixed points, without relying on the detailed form of $J_{ij}$.

For {\it asynchronous} dynamics (sequential or random sequential) and
symmetric $J_{ij}$ the 
distribution of configurations relaxes eventually to the Boltzmann
distribution   eq.(\ref{gibbs2}).

In the noiseless ($T=0$) case, each contribution to the utility is
increased (or remains constant) after each agent $i$ updates her
decision, as can be seen from  
\bear
 U_i(t+1) & = &  h^t_i(t) S_i(t+1) \nonumber\\ 
   & = &    h^t_i(t) \mbox{\rm sgn}(h^t_i(t)) =  |h^t_i(t)|
\nonumber\\ 
   & \ge &   h^t_i(t) S_i(t) = U_i(t) .
\eear
Since the total utility is bounded from above, in absence of asymmetry
the system will asymptotically be driven to a fixed point attractor 
which is either a
local or a global maximum of the utility functional.

While in the case of  symmetric interactions
 there exist only fixed points,
with  asymmetric $J_{ij}$  
cycles of longer length appear.
Eventually in the limit case  of anti-symmetric interactions
($J_{ij} = - J_{ji}$)  Gutfreund, Reger and Young (1987),
have proved that  there exist only 2-cycles.

In the case of  {\it synchronous dynamics} and symmetric interactions
the asymptotic distribution   depends on $\beta$ (and 
consequently is non Gibbsian) and can only formally be written
in the Boltzmann form   (Amit 1989):
 \be
P_e(S) \sim \exp( \beta U_{\beta}(\tilde S)) 
\ee
with 
\be
U_{\beta}({\tilde S}) = 1/\beta \sum_i \ln (2 \cosh \left( \beta  h_i^t)
 \right)  
\ee

In the $T \to 0$ (i.e. $\beta\rightarrow\infty$) limit 
$ U_{\beta}(\tilde S)$ reduces to 
\be
\hat U =  \sum_{j\neq i} J_{ij} S_i(t) S_j(t-1) + h_i  S_i(t)
\ee

The dynamical process now maximizes $\hat U$ (often called the
stability function), since
\bear
\hat U(t+1) & = & \sum_i h^t_i(t) S_i(t+1) \nonumber\\ 
   & = &   \sum_i h^t_i(t) \mbox{\rm sgn}(h^t_i(t)) = \sum_i |h^t_i(t)|
\nonumber\\ 
   & \ge &  \sum_j h^t_j(t) S_j(t-1) = \hat U(t) .
\eear

If $J_{ij}$ is symmetric 
one sees that   
\bear
\Delta \hat U & = & \hat U(t+1) - \hat U(t) \nonumber \\
 & = & \sum_i[S_i(t+1) - S_i(t-1)] \sum_{ j\neq i} J_{ij} S_j(t)
\eear
implying that $\hat U(t+1) \ge \hat U(t)$. The $\hat U(t)$
remains unchanged either when the consecutive states are identical,
i.e. the  system has reached a fixed point, or when the two states
alternate $S_i(t+1)=S_i(t-1)$, i.e. the system has reached a
2-cycle. The existence of 2-cycles with symmetric interactions is a
unique feature of synchronous dynamics.  

If, on the other hand $J_{ij}$ is antisymmetric,  then
$\hat U(t)$ changes by
\bear
\Delta \hat U & = & \hat U(t+1) - \hat U(t) \nonumber \\
 & = & \sum_i[S_i(t+1) + S_i(t-1)] \sum_{ j\neq i} J_{ij} S_j(t)
\eear
and thus, again $\Delta\hat U(t) \ge 0$,  where the equality holds
when $S_i(t+1)=-S_i(t-1)$. Therefore, for antisymmetric interactions the
only attractors are 4-cycles, since the first and the third states are
inverses of each other, as are the second and fourth state.

Eventually  we remind that the  number of fixed points is the same
for sequential, parallel and for random sequential dynamics. 

\section{Model}
\label{Model}

We consider a population of $N$ agents, ordered on a one-dimensional
space and labeled by a variable $w_i$ which represents their position in
the social spectrum and in a broad sense their wealth. Agents' wealth is
chosen randomly, from the uniform distribution in the interval $[0, W_0]$,
and does not change with time.  
In this paper, wealth serves as an index of social status rather than the
source of a budget constraint, as discussed below. A more realistic
situation with consumers arranged over a multidimensional space
(accounting, for example, for differences in  age, education, etc.)
should be considered, but for simplicity we  only 
use one parameter to characterize the  agents, namely their wealth.

According to our previous description, the state of our population 
 $\tilde S(t) = (S_1(t), \ldots, S_N(t))$, evolves according
to a jump Markov process. Time evolves discretely  and at
each step an agent $i$ has a binary choice either to consume one
indivisible unit of a good in which case $S_i=1$, or not to consume, in
which case $S_i = -1$.  
For some product that exists, or appears for the first time in the
market, and given an initial state at time $t=0$, each agent decides
whether to consume or not at every subsequent time step, doing so if this
action provides positive utility. 

For example, a new restaurant opens at time zero and in each subsequent
time period agents decide whether to visit it or not. 

The utility function  $U_i(t)$ is specified in eq.~(\ref{utility}).
The local field $h_i$ characterizes the intrinsic value of the
good to agent $i$. 
This term contains all private factors that affect his consumption
decision. 
A product with $h_i = 0$ for all agents is called a ``fashion'' good,
while a ``status'' good has a positive intrinsic value that might be
well-suited to the characteristics or tastes of a particular class of
consumers.

\begin{figure}
  \epsfxsize 7.5cm  \centerline{\epsffile{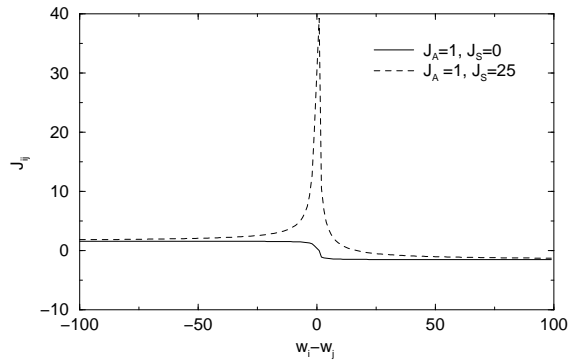}}
\caption{ $J_{ij}$  as a function of $(w_i - w_j)$ for $J_A = 1$ and  $J_S=1,25$.} 
 \label{Pot}
\end{figure}

Each agent interacts with all the others, and the coupling constants
$J_{ij}$ are functions of the agents' status according to:
\be
J_{ij}  =  -J_A \arctan (w_i -w_j) +  J_S \left[ \frac{\pi}{2} - \arctan
|w_i -w_j| \right] 
\label{Jij}
\ee
The coefficients $J_A, J_S$ are taken positive. 
The asymmetric term, proportional to $J_A$, gives a negative
contribution to the utility function $U_i$ if $w_i>w_j$ and a positive
contribution if $w_i<w_j$. This means that agent $i$ wishes to
distinguish  herself from the poorer  while imitating the richer.
The second contribution, proportional to $J_S$, always generates positive
utility and expresses peering effects among consumers of similar status.
The level of asymmetry $k$ is defined by the ratio $k = J_A/J_S$.
Both of these contributions saturate with distance to a constant
value. 
In Fig.~(\ref{Pot}) $J_{ij}$ is plotted as a function of $(w_i - w_j)$.

We shall further assume that agents make expectations about the choice
of others according to
\be
E_i[S_j (t)] = \frac{1}{M} \sum_{m = 1}^{M} \alpha_t S_j(t-m) 
\ee
Various scenarios could be considered by changing the length of the memory
$M$ of each agent and the weights, $\alpha$, that agents put on 
past realizations in the expectation formation process.
We have focused on the case  
\be
\alpha_t =  e^{-\lambda t}
\ee 
with $\lambda > 0$  simulating in this way a fading memory.

Eventually we introduce  a set of parameters $c_i$ which  account
 both for the price of the good and  
for the idiosyncratic costs  that each agent individually may  face.
In our previous example of the restaurant, the cost $c_i$
 for a consumer may be larger (lower) if (s)he lives farther
(closer) to it.
We assume that costs  act  merely as thresholds and the decision of
agent $i$ on whether to buy the good or not depends only on whether his 
utility of buying one unit $U_i$ is positive and larger than $c_i$. 
This means that in eq.(\ref{HeatBath},  \ref{parallel}, \ref{sequential})   
\begin{equation}
h_i^t \rightarrow h_i^t - c_i.
\end{equation}
Nonetheless  all agents are assumed to
possess sufficient liquidity at all points in  time and wealth
constraints are never binding.

\section{Simulations and Results}
\label{results}

We have performed computer simulations to study the model in
eq.~(23, \ref{Jij}, 39) for each of the evolution algorithms of section
\ref{evolution}.  We studied both the time
evolution of  total consumption as well as the spatial
distribution of consumption across the social spectrum. 
We explored the various patterns generated
with different choices of the parameters. 
Our plots refer to  lattice size $N=800$.
We have taken the agents'
wealth to be uniformly distributed in the interval $(0, W_0)$ with
$W_0=100$. 

\vskip 0.3cm
\noindent{\bf (a) $M=1, T=0, G=0, C=0$}
\vskip 0.3cm

Initially we study  the case of a fashion good which has no  intrinsic value
for the consumers, i.e. $h_i=0$.

In analogy with Spin Glass models (Mezard et al. 1987) 
with asymmetric interaction\footnote{Our  model could be considered as a 
particular case of the asymmetric  SK model (Iori and Marinari 1997) 
with  the couplings taken
according to eq.~(\ref{Jij})  instead of being chosen randomly.} 
the nature of the attractors  depends in a complex manner
on the level of asymmetry $k$.
Given the analogy between the two systems  we  briefly resume in the
following some of the features of the SK model. N\"utzel and Krey (1993) 
found that only fixed points or periodic attractors with period two are present
in the nearly symmetric SK model  while longer attractors appear in the
case of  highly asymmetric  couplings
 (they locate the  transition  at $k_c \sim 1/\sqrt{3}$).
While in the nearly symmetric case the average length of the attractor
$<l>$ (measured by the number of different configurations the system
goes through before repeating an identical  sequence) is independent of
the system size, $N$,  the  dependence of   $<l>$ is exponential in $N$
for highly  asymmetric couplings.
Moreover, very long transients are present  and 
 the typical number of updates before the system relaxes to the
attractor grows as a power of $N$ at low asymmetry 
and   is exponential in $N$ for high asymmetry.
Furthermore the transient exhibits chaotic behaviour, 
i.e. sensitive dependence on the initial conditions (Crisanti et al. 1993).

In our model  we do not observe the long transients before the
system reaches the attractor,  neither at high or low asymmetry.
Nonetheless we found an interesting dependence of $<l>$ on $k$
which we  report in the following.

\begin{figure}[htb]
  \epsfxsize 7.5cm \centerline{\epsffile{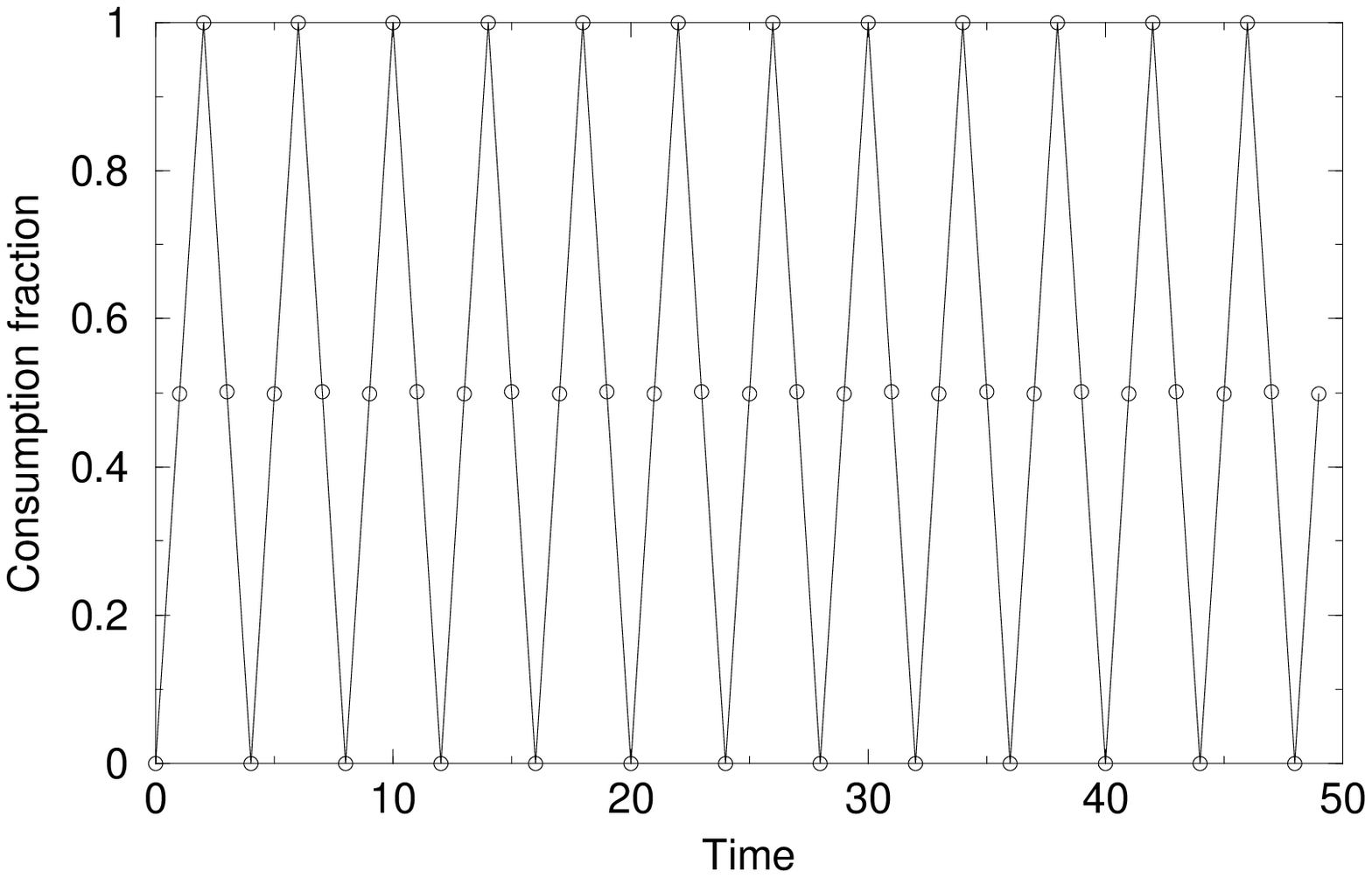}}
  \epsfxsize 7.5cm \centerline{\epsffile{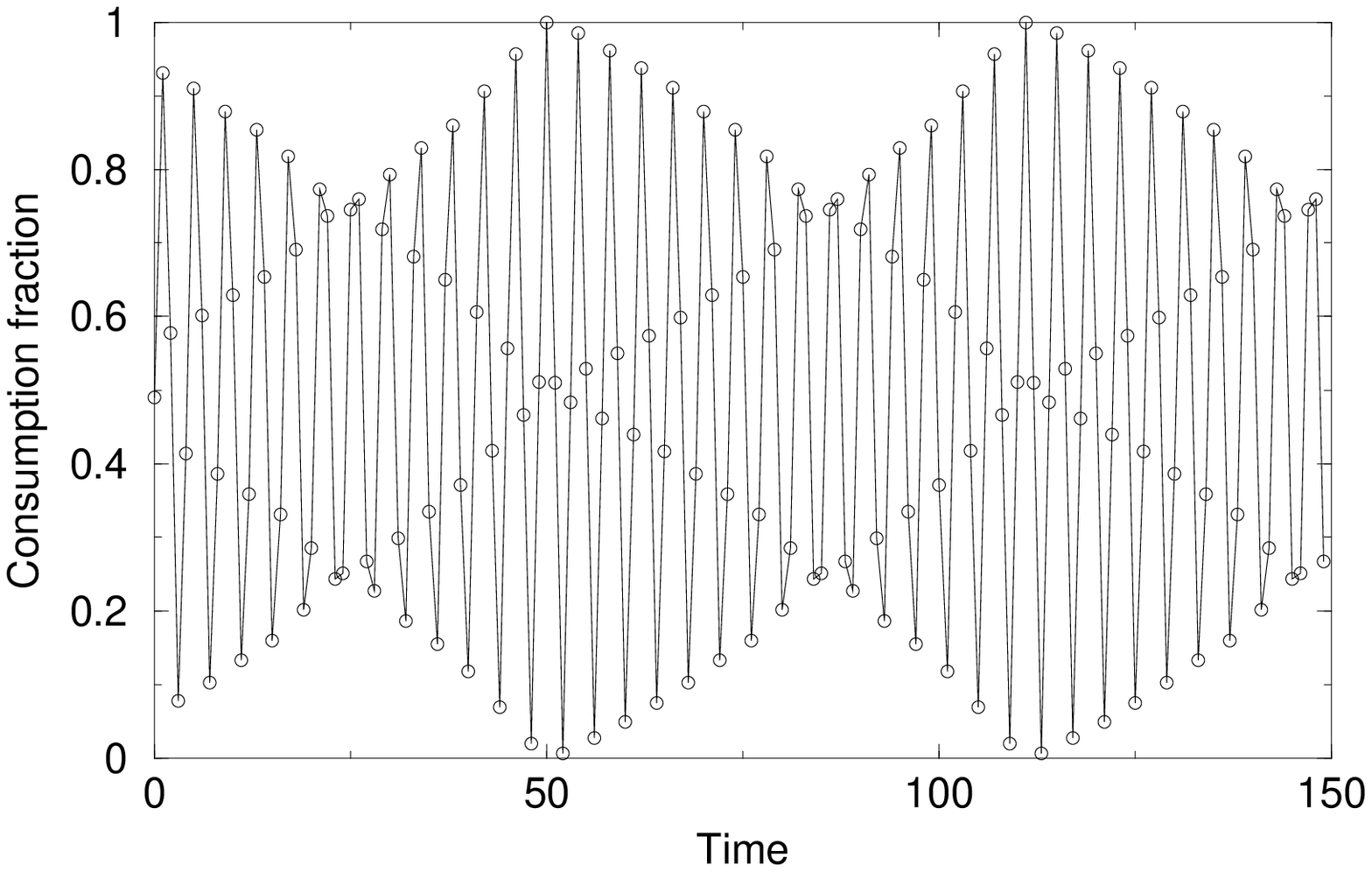}}
  \epsfxsize 7.5cm \centerline{\epsffile{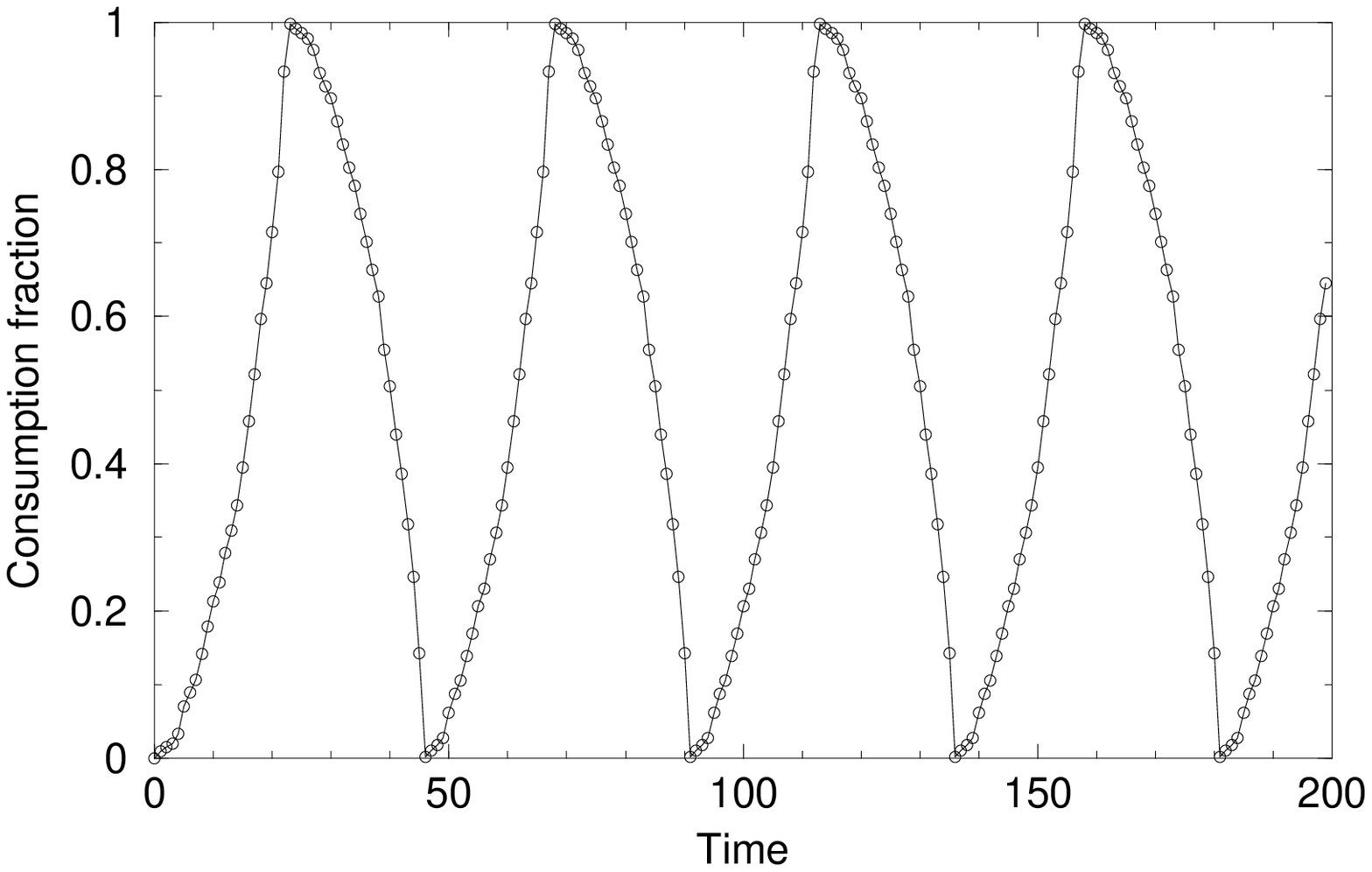}}
\caption{Total consumption with parallel dynamics as a function of time
for $M=1, T=0, G=0, C=0$
and, from top to bottom, (a) $k = \infty$, (b) $k = 3$, (c) $k = 1/25$.}  
 \label{fig2}
\end{figure}

\begin{figure}[htb]
  \epsfxsize 8cm  \centerline{\epsffile{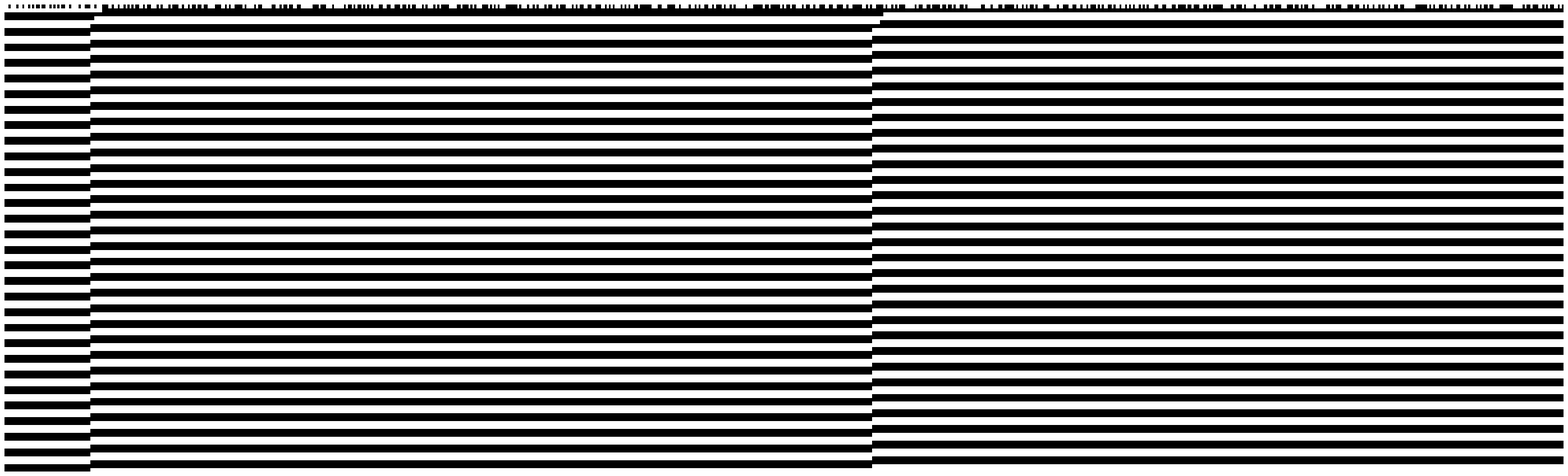}}
\vskip .2cm
  \epsfxsize 8cm  \centerline{\epsffile{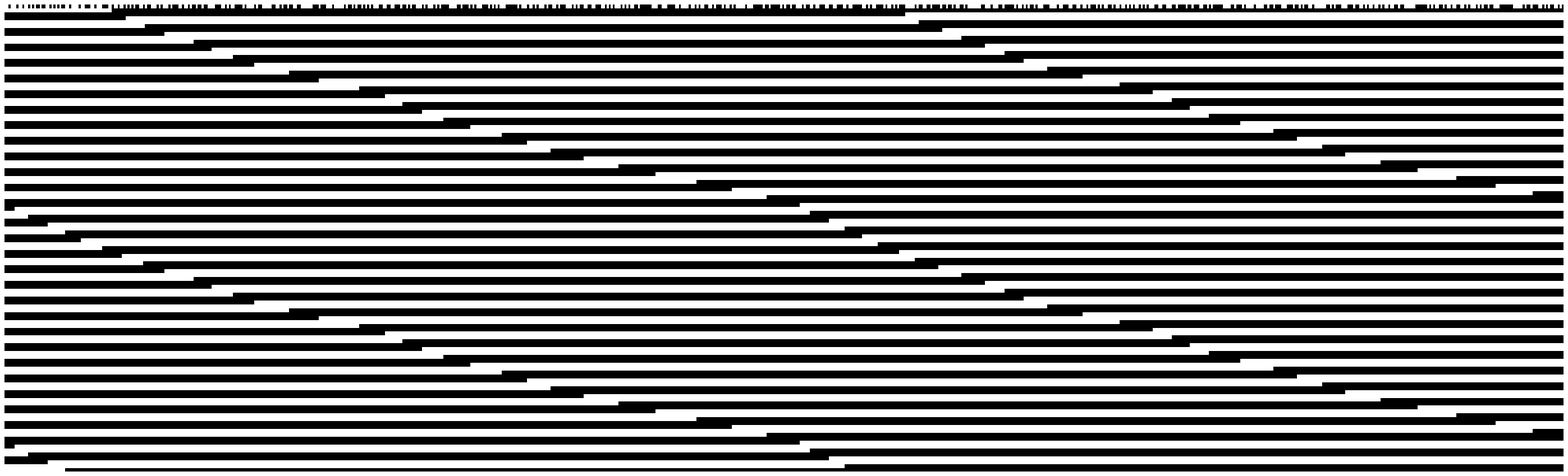}}
\vskip .2cm
  \epsfxsize 8cm  \centerline{\epsffile{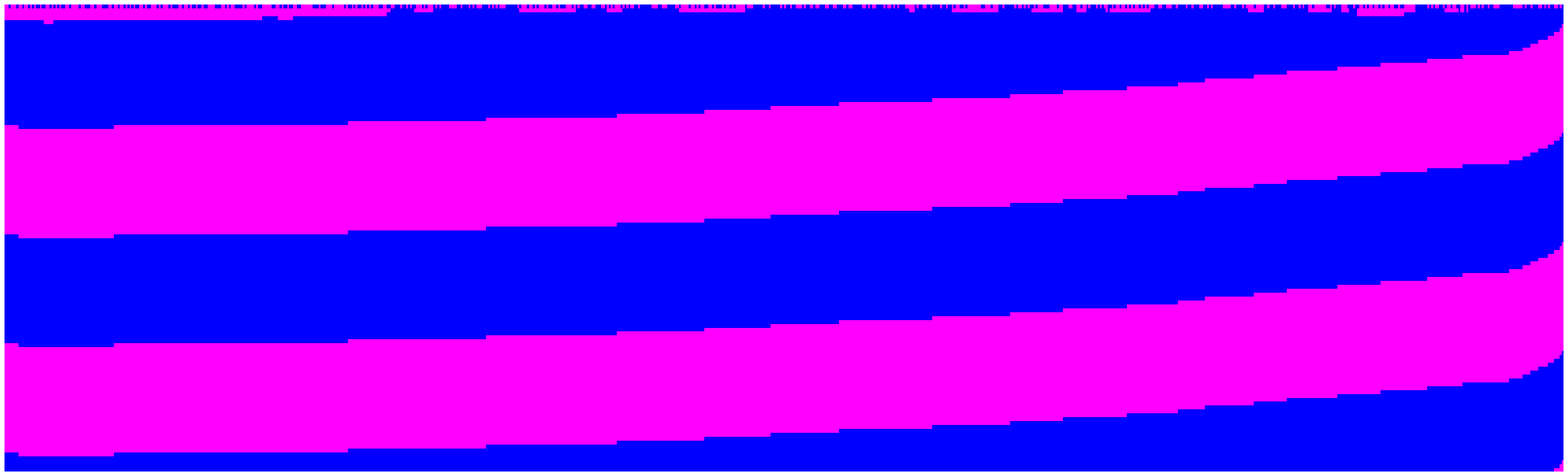}}
\vskip .4cm
\caption{Wave dynamics  with parallel dynamics for $M=1, T=0, G=0, C=0$
and top (a): $k=\infty$, center (b): $k=1/4$,  
bottom (c): $k=1/25$. The dark
color corresponds to those agents consuming and light color to those not
consuming. Wealth is increasing from left to right along the horizontal
axis. The vertical axis represents time (here $t=120$ steps) which
increases from top to bottom.} 
 \label{fig3}
\end{figure}

\begin{figure}[tb]
\epsfxsize 8cm \centerline{\epsffile{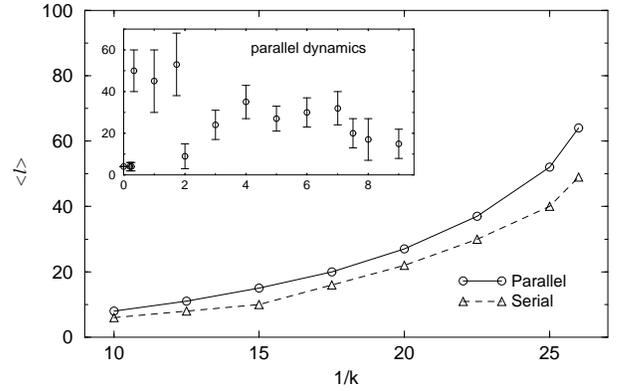}}
\caption{Length of the attractor for  $M=1, C=0$, $G=0, T=0$ as a
 function of $1/k$. We expand  the region at large  $k$ in the inset for
parallel dynamics. The main figure compare parallel dynamics (circles)
with sequential dynamics (triangles) in the low asymmetry limit.}
 \label{fig4}
\end{figure}

For  parallel dynamics and  $k \rightarrow \infty$,  as we anticipated
in the previous  section, the system only exhibits 4-cycles
(fig.~(2a)) and  consumption propagates very fast through the social spectrum 
(fig.~(3a)). 
Note that in fig.~(3) time increases from top to bottom   
and it is the rich who start consuming first in order to
distinguish themselves from the poor. 

The average length of the attractors as a function of $1/k$ is plotted in
 fig.(\ref{fig4}).
The behaviour of the system is  very different in the two 
regions of high and low asymmetry.   
First of all we notice that in the large $k$ limit the length of
the attractor varies a lot from configuration to configuration 
(which explains the large error bars in the inset of fig.~(\ref{fig4}))
while at low asymmetry the length is practically  independent from the
realization of the $J_{ij}$ (we also checked that  in this regime the
length of the attractor is almost independent from the system  size).

The spatial distribution of  consumption also differs in the two regions.
At  high asymmetry, as shown in fig.~(3b), consumption propagates across
the social spectrum  very fast. The difference with respect to the
previous case is that now the location (in $w$) where agents stop/start
consuming  is different at different times and this  generates a
modulating effect on the total  consumption (fig.~(2b)).
On the other hand, at low asymmetry, consumption propagatess slowly and
waves of longer period appear  as shown  in fig.~(3c).
We locate the  transition  at $k_1 \sim 0.1$.
 As $k$ is decreased  further, the period of the waves (i.e. the length
of the attractor) increases, as can be observed from the right end side
of  fig.(\ref{fig4}).  
As $k$ decreases  below a certain value ($k_2 \sim 1/27$),
waves disappear and the system is attracted towards a fixed point. The
fixed point is characterized by  either all agents  consuming or nobody
consuming.\footnote{These two configurations have the same utility
 as the transformation $S_i\rightarrow -S_i$ is a
symmetry of the system. The initial condition determines towards
 which of the two states the system relax.} 

Note that in fig.~(\ref{fig3}c) at  the beginning of each 
cycle consumption propagates rather slowly but
it spreads faster as it moves down to the poorer. $S$-shaped curves,
similar to those in fig.(2) have been observed in the case of diffusion of
innovation (Rogers 1995).

\begin{figure}[hb]
  \epsfxsize 8cm  \centerline{\epsffile{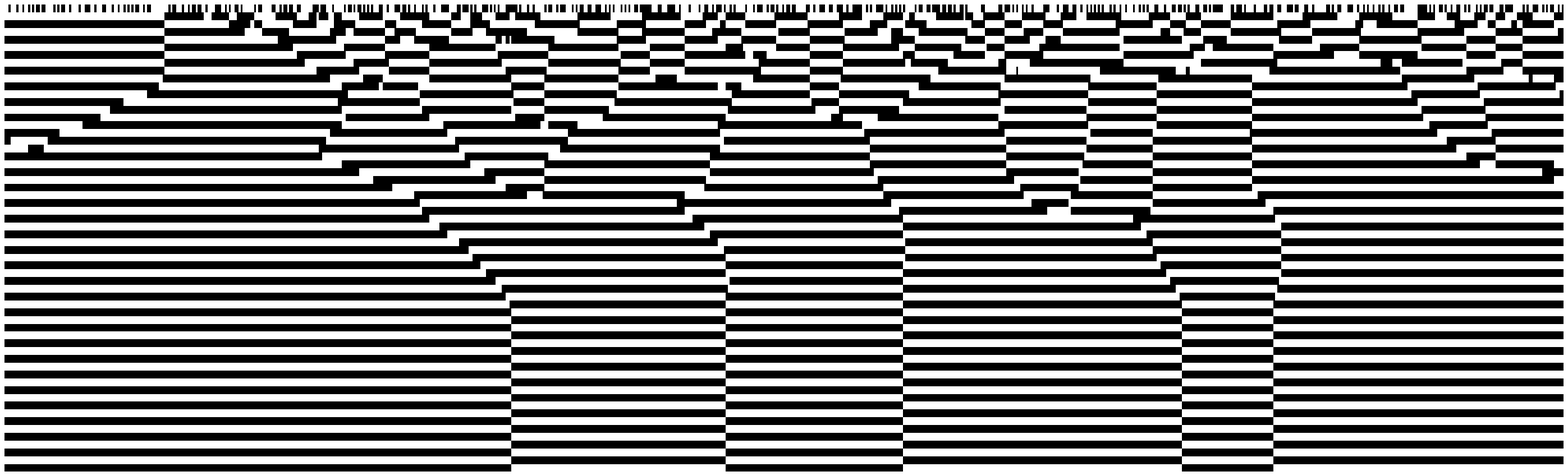}}
\vskip 0.5cm
\caption{Consumption behaviour with sequential dynamics when
$k=\infty$,  $M=1, T=0, G=0, C=0$.}
 \label{fig5}
\end{figure}

For sequential dynamics and  $k \rightarrow \infty$,
 as anticipated,
 in the previous  section, the system only exhibits  2-cycles
 as depicted in fig.~(\ref{fig5}). 
Notice that after a short transient period during which
consumption is irregular, the system ends into a periodic attractor
characterized by  agents clustering into  groups that synchronously alternate
 their consumption behaviour. Starting with different initial conditions 
 affects the position and the number of clusters
which form.
As we reduce $k$ towards the low asymmetru region the dynamics of
sequential updating  looks similar to that of parallel dynamics
although  the length of the attractors 
is shorter  as can be seen from fig.(\ref{fig4}). 

For  random sequential dynamics periodic attractors do not exist.
Nonetheless in the low asymmetry region waves propagate through the
 system in a fashion similar to parallel and sequential dynamics.
At very low  $k$ again the dynamics converges to a fixed point
where either all or nobody is consuming. 

\begin{figure}[thb]
\epsfxsize 8cm \centerline{\epsffile{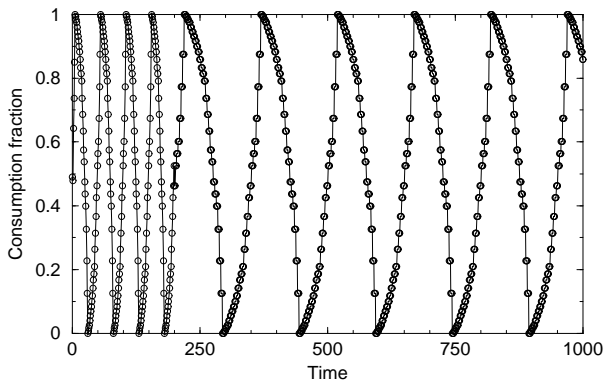}}
\caption{Total consumption with $k=1/25, C=0$, $G=0, T=0$, and  $M=1$ for
time $t<200$ and $M=5$ for $t>200$.}
 \label{fig6}
\end{figure}

\vskip 0.3cm
\noindent{\bf (b) $M > 1, T = 0, G = 0, C = 0$}
\vskip 0.3cm

\begin{figure}[bh]
\epsfxsize 8cm \centerline{\epsffile{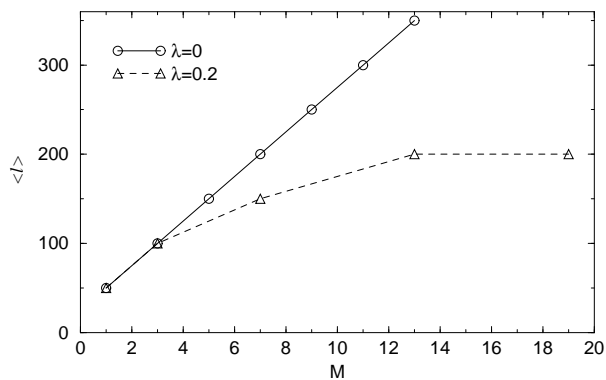}}
\caption{Length of the attractor for  $k=1/25, C=0$, $G=0, T=0$
increasing $M$.  Two cases are compared $\lambda=0$ (circles) and
$\lambda=0.2$ (triangles).}
 \label{fig7}
\end{figure}

Adding  finite memory to the system does not destroy the waves but  
it changes their frequency and  the corresponding length of the attractor. 
This can be inspected from fig.(\ref{fig6}) where  we have 
run the simulation with memory  $M=1$ for the initial  200 steps
and then continued from there on with $M=5$. Fig.(\ref{fig7}) shows how 
the length of the attractor increases with $M$, with all other parameters fixed.
In fig.~(7) we compare  two cases, one with $\lambda=0$ which
corresponds to a non-fading memory, and another with $\lambda=0.2$,
which corresponds to a slowly fading one. 
While at $\lambda=0$ the dependence of $<l>$ on $<M>$ is linear, 
for  $\lambda=0.2$ we observe that the length of the attractor increases
at a slower rate  and eventually saturates (when $t$ is sufficietly large and
$\alpha_t \sim 0$).

\begin{figure}[htb]
\epsfxsize 8cm \centerline{\epsffile{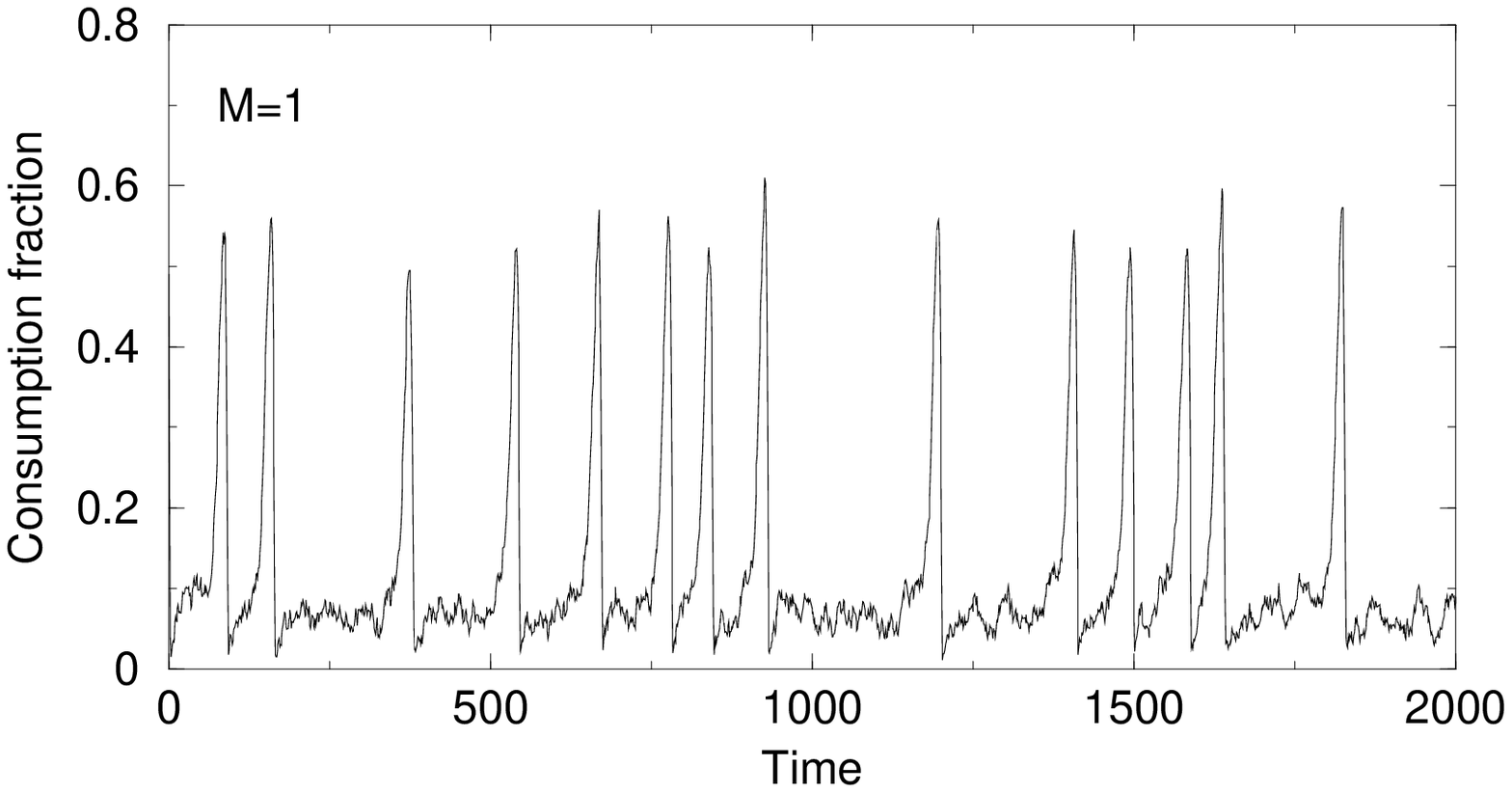}}
\epsfxsize 8cm \centerline{\epsffile{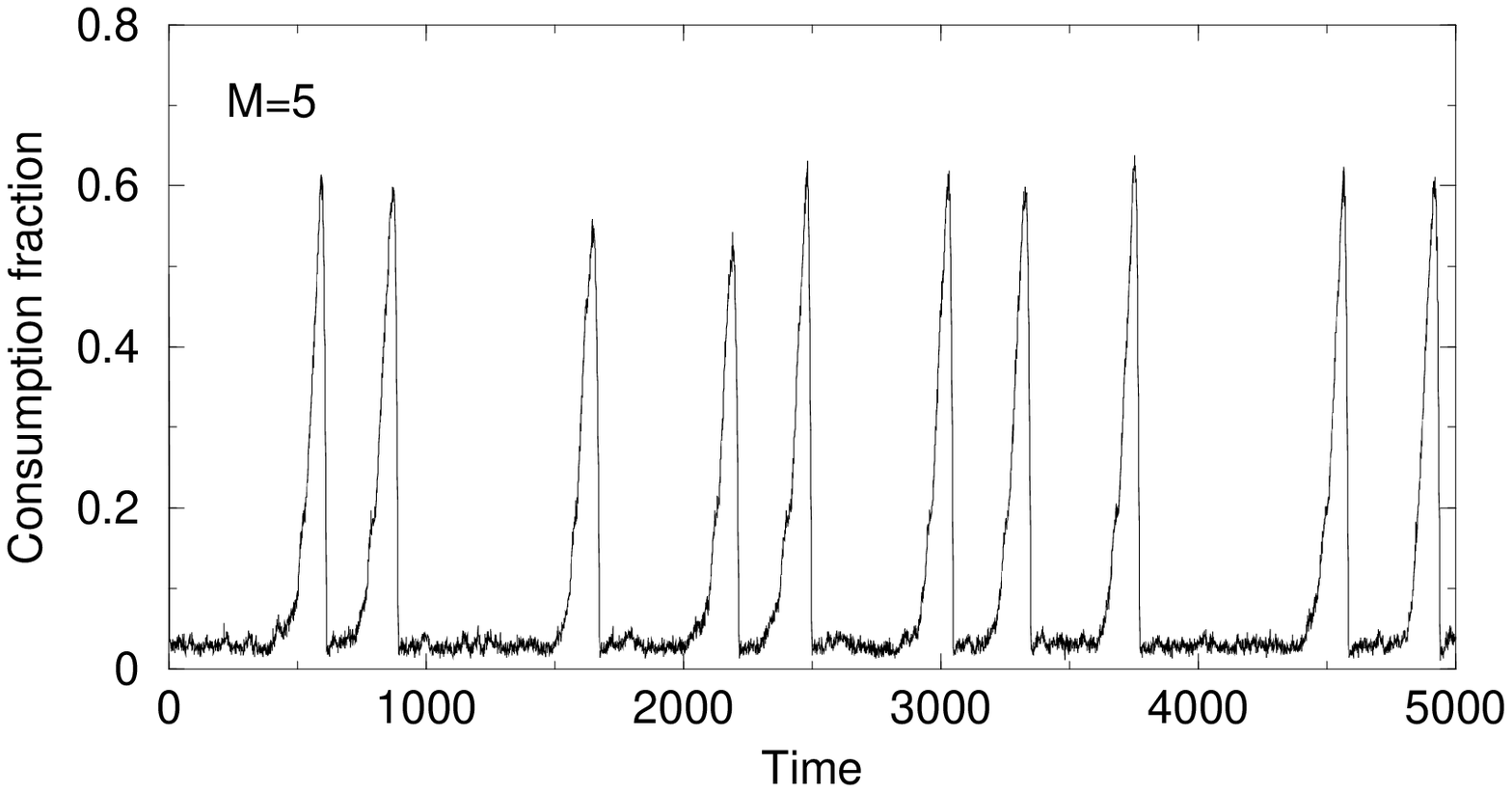}}
\caption{Waves in consumption with $k=1/25, C=1.3$, $G=0, 
T=1$ and (a) $M=1$ (top), (b) $M=5$ (bottom). }
 \label{fig8}
\end{figure}

\begin{figure}[ht]
  \epsfxsize 8cm  \centerline{\epsffile{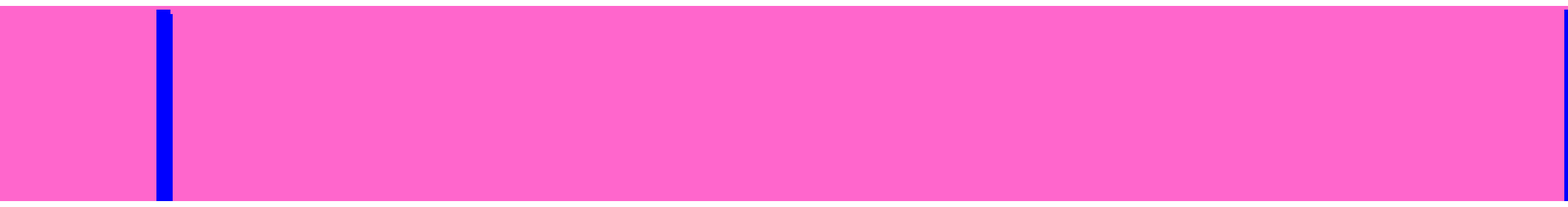}}
\vskip 0.5cm
  \epsfxsize 8cm  \centerline{\epsffile{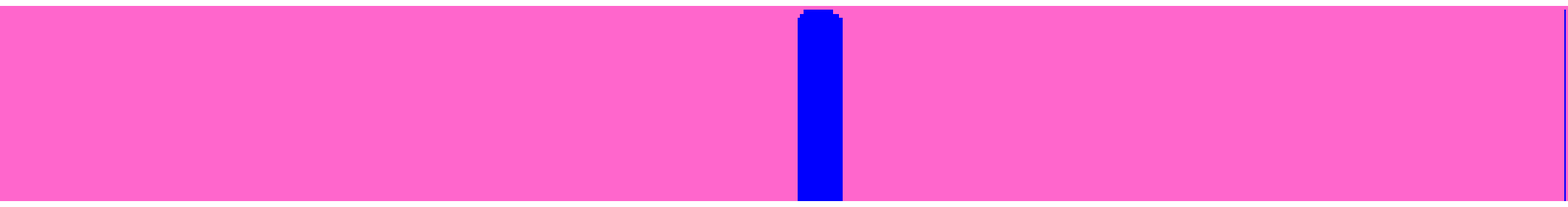}}
\vskip 0.5cm
  \epsfxsize 8cm  \centerline{\epsffile{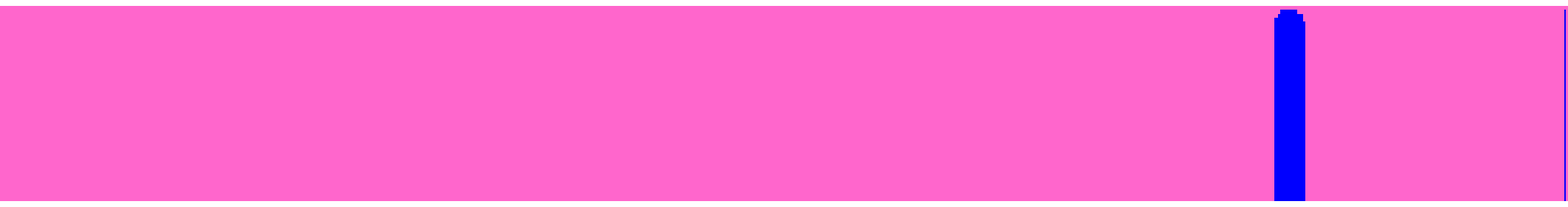}}
\vskip 0.5cm
  \epsfxsize 8cm  \centerline{\epsffile{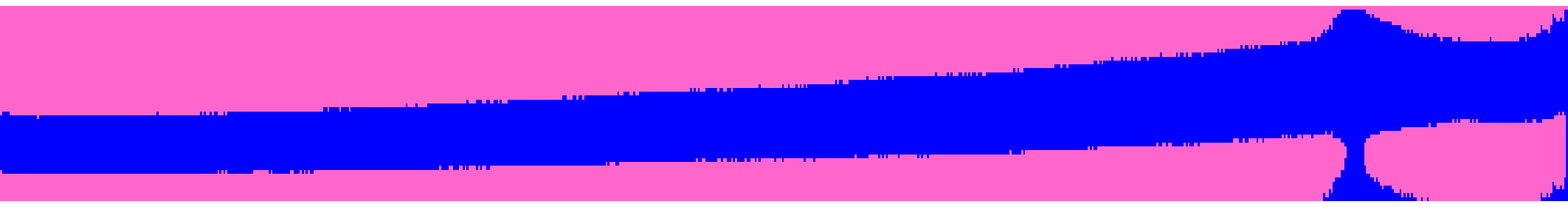}}
\vskip 0.5cm
\caption{Consumption behaviour with parallel dynamics when
$k=1/25$, $C=0.5$, $G=1$ $T=0$, $M=1$, and $w_m = 10, 50, 80, 85$ (from top to
bottom). Dark (light) color is for those consuming (not consuming). Only
when the good is suitable to the consumers located  at the top of the
social scale a consumption wave propagates. In all other case the good
enters and finds a stable niche.} 
 \label{fig9}
\end{figure}

\vskip 0.3cm
\noindent{\bf (c) $M \geq 1, T > 0, G = 0, C > 0$}
\vskip 0.3cm

We now examne the case of $G=0$ and non-zero costs.
Costs are  chosen randomly  for each agent  from the
uniform distribution in the interval $(-C,0)$ and are fixed in time.

Averaging over many  realizations of agent's wealth we find the
average  critical value $C_c=0.58 \pm 0.03$ above which, at $T=0$, 
nobody consumes. 
Keeping the same values for all  the other parameters as before but 
taking  $C=1.3$ (which is well above $C_c$) we added noise. The effect of
noise is that it induces a certain number of agents  to  consume despite
the high costs.  
If the number of  agents who consume  exceeds a critical mass then
consumption waves, of varying amplitude, emerge spontaneously, even 
 though at irregular time intervals  (see Fig.~(\ref{fig8})).
Adding memory has the effect to make this occasional consumption waves
more sparse.  If $T$ becomes  much  larger, waves disappear  and
agents consume randomly.

\vskip 0.3cm
\noindent{\bf (d) $M=1, T=0, G > 0, C >0$}
\vskip 0.3cm

We now consider the case of a ``good'' for which agents manifest opposite  preferences 
independently of their social status, i.e. the intrinsic value of the good $h_i$ 
is a random number, constant in time,  chosen for each agent $i$
independently from the interval $(-G,G)$.  

For $G$ smaller that a certain value $G_c$ 
we still found periodic attractors with the length of the
attractors  decreasing  with $G$ increasing. 
Nonetheless   we do not observe consumption waves
propagating along the social spectrum unless $G$ is very small.
On the other hand if agents  have strong
preferences, the contribution of their private utility dominates $U_i$ in
eq.(\ref{utility}), and only a  fraction of agents update their decision under
the influence of the  others. 
Eventually, when increasing $G$  above $G_c$  the dynamics
converges to a fixed point. Nonetheless the fixed point is not only
dictated  by the private utility; the social component modifies the
natural distribution, i.e. the one where $S_i$ has the same sign of $h_i$.  

Another interesting case is to consider a ``status'' good which is 
designed  to meet the needs of  a specific group of consumers.
In this case we assume the good mainly  provides individual  utility
to consumers whose wealth is distributed around a given value $w_m$ choosing:
\begin{equation}
h_i=h(w_i) = \frac{G}{(w_i - w_m)^2} 
\end{equation}

To analyze  the interplay between the intrinsic value of the good
(here $G$) and the costs, we fix $C$ to be larger than $C_c$
so that waves  do not emerge in the case of a fashion good ($G=0$). 
Therefore  it is only the intrinsic value of the good which can trigger 
consumption.

The results which follow refer to the case  $k=1/25, G=1, T=0, M=1$
and $C=0.5$.
Different behaviours are found, depending on the position of the 
maximum ($w_m$) of $h(w_i)$.
For $w_m < 85$, (we remind that the agents' wealth is distributed
between zero and $W_0 = 100$),   the good enters  the social spectrum
around $w_m$, possibly migrates through the closest social classes and
then finds a stable niche (see top  three cases in Fig.~(\ref{fig9})). 
Only  when $w_m \sim W_0$
 waves emerge and spread throughout the whole social spectrum
(bottom case in  Fig.~(\ref{fig9})).

\section{Conclusions}
\label{conclusions}

In this paper we have focused on a potentially important mechanism that
drives consumption decision: the interaction among heterogeneous
consumers. Particular attention has been paid to the role of the
asymmetry of interactions  and the dynamical updating  rules.
In the sociology literature, interactions among individuals, belonging
to similar or different social circles, are often seen as a major
mechanism  that determines new styles of behaviour.
We studied how peering, distinction and aspiration effects,
in addition to the intrinsic values of a good, generate
different  consumption patterns, under the assumption that  information
about the consumption behaviour of  agents is public (we imagined that
each agent knows the past behaviour of all other agents). 
Nonetheless collective behaviour may be affected by the structure of the
communication channels. 
To check the sensitivity of our results with respect to the 
size of the reference group of each agent
we  have examined  the case where individuals only communicate
with a subset, chosen at random,  of the entire population.
If the size of the subsets is  as small as $5\%$ of the total population
we still observe (in the low asymmetry limit) waves propagating through
the system whose frequencies increase when $k$ decreases.

\section*{Acknowledgments}
We are grateful to C. Hiemstra and S. Jafarey for helpful comments.
V.K. also wishes to thank
the University of Essex for the kind hospitality provided during the
initial stages of this work.


\section*{References}
\noindent
Akerlof G. A., {\em Social Distance and Social decision, Econometrica},
Vol. 65, No. 5 (1997), 1005-1027. 
\\
\\
Albin P.S., {\em Barriers and bound to rationality}, 
Princeton University Press (1998)
\\
\\
Amit D., {\em ``Modelling Brain Functions''}, Cambridge University Press (1989).
\\
\\
Anderson S., A. de Palma, and Thisse J.-F., {\em ``Discrete Choice Theory
of Product Differentiation''}, MIT press, (1992).
\\
\\
 Aoki M., {\em ``New approaches to macroeconomic modeling''},
Cambridge University Press (1996). 
\\
\\
 Axelrod R., {\em The complexity of cooperation}, Princeton University Press (1997)
\\
\\
Blume L., {\em ``The Statistical Mechanics of Strategic Interaction''},
Games and Economic Behaviour, 5, (1993), 387-423.
\\
\\
Benabou R., {\em ``Equity and Efficiency in Human Capital Investment: The
Local Connection''}, Review of Economic Studies, 62 (1996) 237-264.
\\
\\
Brock W., {\em ``Asset Price Behaviour in Complex Environments''}, Mimeo,
Dept. of Economics, University of Wisconsin, Madison (1995). 
\\
\\
Brock W. and Durlauf S., {\em ``Discrete choice with social interactions''},
Mimeo, Dept. of Economics, University of Wisconsin, Madison (1995).
\\
\\
Chwe M.~S.-Y., {\em ``Communication and Coordination in Social Networks''},
Review of Economic Studies, 67, (2000), 1-16.
\\
\\
 Corneo G. and  Jeanne O.,
{\em ``A Theory of Fashion Based on Segmented Communication''}, Mimeo, Dept.
of Economics, University of Bonn (1994).
\\
\\
 Cowan~R.,~Cowan~W.~and~Swan~P.,
 {\em ``Waves in Consumption with Interdependence among Consumers''}, 
 MERIT preprint,  Research Memoranda series, N. 011.
\\
\\
A. Crisanti, M. Falcioni and A. Vulpiani,
{\em  Transition from regular to complex behaviour in a discrete
 deterministic asymmetric neural network model}, J.Phys. A26 (1993) 3441-3454.
\\
\\
Durlauf S., {\em  ``Statistical mechanics approaches to socioeconomic behavior''},
in The economy as an evolving complex system II, W.B. Arthur,
S.N. Durlauf and D. Lane, eds., Redwood City: Addison-Wesley (1997).
\\
\\
Durlauf, S., {\em ``A Theory of Persistent Income Inequality''}, Journal of
Economic Growth, 1, (1996), 75-93.
\\
\\
F\"ollmer, H., {\em ``Random Economies with Many Interacting Agents''},
J. of Mathematical Economics, 1, (1974) 51-62.
\\
\\
S. Galam, {\em ``Spontaneous coalition forming: a model from spin glass''},
preprint, http://xxx.lanl.gov/abs/cond-mat/9901022.
\\
\\
R.J. Glauber, {\em ``Time-dependent statistics of the Ising model''}, 
J. Math Phys. 4, 294 (1963).
\\
\\
Hammersley J. M. and Handscomb D.C., {\em ``Monte Carlo Methods''}, Methuen's
Monographs, London (1975).
\\
\\
G. Iori and E. Marinari,  {\em On the Stability of the Mean-Field 
Spin Glass Broken Phase under Non-Hamiltonian  Perturbations}, 
J. Phys. A: Math. Gen.   {\bf 30} (1997) 4489-4511.
\\
\\
Kauffman S. A., Journal of Theoretical Biology, 22 (1969) 437.
\\
\\
 Kirman A., {\em ``Economies with interacting agents''},
in The economy as an evolving complex system II, W.B. Arthur,
S.N. Durlauf and D. Lane, eds., Redwood City: Addison-Wesley (1997).
\\
\\
Mezard M., Parisi G. and Virasoro M.A., {\em ``Spin glass theory and
beyond''}, (World Scientific, Singapore 1987).
\\
\\
Morris S., {\em ``Contagion''}, Review of Economic Studies, 67, (2000), 57-78.
\\
\\
M\"uller B., Reinhardt J., Strickland M.T., 
{\em ``Neural Networks : An Introduction''}, Springer-Verlag, 1995.
\\
\\
N\"utzel K. and Krey U., 
{\em ``Subtle dynamic behaviour of finite-size Sherrington-Kirkpatrick
spin glasses with non-symmetric couplings''}, J. Phys. A: Math. Gen. 26
(1993) 591-597.
\\
\\
Rogers E.M., {\em ``Diffusion of Innovations''}, The Free Press, New York, 1995.
\\
\\
Samuelson L., {\em ``Evolutionary Games and Equilibrium Selection''}, MIT
Press, 1997.
\\
\\
Sargent T. J., {\em ``Bounded Rationality in Macroeconomics''}, Clarendon
Press, 1994.

\end{document}